\documentclass[12pt]{article}
\usepackage{calrsfs}
\usepackage{calligra}
\usepackage{amsmath,amssymb}
\usepackage{graphics,graphicx}
\def\a{\alpha}
\def\b{\beta}

\def\d{\delta}
\def\e{\epsilon}

\def\g{\gamma}

\def\k{\kappa}

\def\t{\theta}
\def\r{\rho}
\def\s{\sigma}

\def\G{\Gamma}

\def\L{\Lambda}
\def\O{\Omega}

\def\ve{\varepsilon}

\def\be{\begin{equation}}
\def\ee{\end{equation}}
\def\arr{\begin{array}{rll}}
\def\ea{\end{array}}
\def\bea{\begin{eqnarray}}
\def\eea{\end{eqnarray}}

\def\N2{$N{=}2$}

\def\>{\rangle}
\def\<{\langle}
\def\+{\dagger}
\def\={\ =\ }

\begin{document}
\vspace{0.5cm}
\renewcommand{\thefootnote}{\fnsymbol{footnote}}
\renewcommand{\thefootnote}{\fnsymbol{footnote}}
\begin{titlepage}
\setcounter{page}{0}
\vskip 1cm
\begin{center}
{\LARGE\bf Super 0--brane action on the coset space  }\\
\vskip 0.5cm
{\LARGE\bf of $D(2,1;\a)$  supergroup }\\
\vskip 1cm
$
\textrm{\Large Dmitry Chernyavsky \ }
$
\vskip 0.7cm
{\it
Laboratory of Mathematical Physics, Tomsk Polytechnic University, \\
634050 Tomsk, Lenin Ave. 30, Russian Federation} \\
{E-mail: chernyavsky@tpu.ru}

\end{center}
\vskip 1cm
\begin{abstract} \noindent
The super 0--brane action on the coset space of $D(2,1;\a)$ supergroup is constructed which involves a set of parameters related by $\kappa$-symmetry.  It describes a massive superparticle propagating near the horizon of the extreme Reissner--Nordstr\"om--AdS--dS black hole which is linked to the recently constructed $D(2,1;\a)$--superparticle [JHEP 1703 (2017) 054] by a canonical transformation.
\end{abstract}

\vskip 1,5cm

\noindent
Keywords: $D(2,1;\a)$ supergroup,  Reissner--Nordstr\"om--AdS--dS black hole, super 0-brane

\end{titlepage}

\renewcommand{\thefootnote}{\arabic{footnote}}
\setcounter{footnote}0

\noindent
{\bf 1. Introduction}\\

\noindent
Superconformal mechanics based upon one or another supersymmetric extension of the conformal group in one dimension $SO(2,1)$ is of interest for several reasons. Classification of supermultiplets of $d=1$, $N$--extended supersymmetry and their interactions is important for understanding supersymmetry in divers dimensions \cite{IKL2,BIKL1,Krivonos_lectures}. In particular, some $d=1$ supermultiplets cannot be obtained by dimensional reduction from higher dimensions. Superconformal models in $d=1$ provide a non--trivial example of the $AdS_2/CFT_1$--correspondence \cite{RJ}. Some models of (super)conformal mechanics are relevant for microscopic description of extreme black holes in the near horizon limit \cite{GibbonsTownsend} and for understanding the Banados--Silk--West effect \cite{AG}.

In the chain of $N$--extended superconformal models the instance of $N=4$ is of particular interest due to the proposal in \cite{GibbonsTownsend} that
an $N=4$ superconformal Calogero model may provide a microscopic description of the extreme Reissner–-Nordstr\"om black hole in the near horizon limit. A plenty of such models have been constructed and investigated in
\cite{W}--\cite{FI} which all relied upon the $d=1$, $N=4$ superconformal group $SU(1,1|2)$. A related line of research concerns the study of (super)conformal particles propagating on near horizon black hole backgrounds \cite{Kallosh}--\cite{Gal} and the construction of novel (super)integrable systems associated with such geometries \cite{BNY}--\cite{HD}. Worth mentioning also is the resent study of the $AdS$ superparticles within the group--theoretic approach \cite{HHJM,HJ,HJM}.

It should be mentioned that $SU(1,1|2)$ is a particular instance of the most general $d=1$, $N=4$ superconformal group $D(2,1;\a)$ which arises at $\a=-1$. Although $D(2,1;\a)$--superconformal mechanics has been extensively studied in the past \cite{IKL2}, \cite{ikl}--\cite{FI3}, a link to the near horizon black hole geometries has been established only quite recently \cite{Galajinsky_D}. In particular, a canonical transformation which relates $D(2,1;\alpha)$--superconformal mechanics based upon supermultiplets of the type
$(3,4,1)$ and $(4,4,0)$ to $BPS$--superparticles propagating near the horizon of the extreme Reissner--Nordstr\"om--AdS--dS black hole in four and five dimensions has been found.
Interestingly enough, the group parameter $\alpha$ was linked to the cosmological constant. Since the results in \cite{Galajinsky_D} were obtained within the Hamiltonian formalism, the natural question arises whether the $D(2,1;\a)$--superparticle in \cite{Galajinsky_D} is dynamically equivalent to the conventional $\kappa$--symmetric super 0-brane propagating on the near horizon Reissner--Nordstr\"om--AdS--dS black hole background.
Note that a similar question for the case of the vanishing cosmological constant remained open for quite some time \cite{Zhou,KZ,bgik}.

The goal of this work is to construct the super 0-brane action on the coset space of the $D(2,1;\a)$ supergroup and to demonstrate that its bosonic part can be linked to a massive charged particle propagating in the near horizon region of the extreme Reissner--Nordstr\"om--AdS--dS black hole.

The work is organized as follows. In Sect. 2 we use the Maurer--Cartan one--forms associated with $D(2,1;\a)$ so as to construct an invariant action on the coset superspace. In Sect. 3 we discuss the gauge symmetry transformations. Sect. 4 is focused on the global supersymmetry of the gauge fixed action. The bosonic part of the action and its physical meaning are discussed in Sect. 5. In Sect. 6 a canonical transformation is constructed which links the $\kappa$--symmetric super 0-brane propagating on the near horizon Reissner--Nordstr\"om--AdS--dS black hole background to the $D(2,1;\a)$--superparticle in \cite{Galajinsky_D}. In the concluding Sect. 7 we summarize our results. Our spinor conventions are gather in Appendix A. Some technical details regarding the explicit form of the global supersymmetry transformations are given in Appendix B.

\vspace{0.5cm}

\noindent
{\bf 2. $D(2,1;\a)$ and invariant action on the coset space}\\

Our starting point is the Lie superalgebra associated with $D(2,1;\a)$ in the notation of \cite{Galajinsky_D}
\begin{align}\label{algebra}
&
[ H,D ]=H\ , && [ H,K ]=2D\ ,
\nonumber\\[2pt]
&
[D,K]=K\ , && [ \mathcal{J}_a,\mathcal{J}_b ]=\epsilon_{abc} \mathcal{J}_c\ ,
\nonumber\\[2pt]
&
\{ Q_\a, \bar Q^\b \}=-2 i H {\d_\a}^\b, &&
\{ Q_\a, \bar S^\b \}=-2\alpha {{(\s_a)}_\a}^\b \mathcal{J}_a+2iD {\d_\a}^\b+2(1+\alpha)I_3 {\d_\a}^\b,
\nonumber\\[2pt]
&
\{ S_\a, \bar S^\b \}=-2i K {\d_\a}^\b, &&
\{ \bar Q^\a, S_\b \}=2\alpha{{(\s_a)}_\b}^\a \mathcal{J}_a+2iD {\d_\b}^\a-2(1+\alpha)I_3 {\d_\b}^\a,
\nonumber\\[2pt]
&
\{ Q_\a, S_\b \}=2i (1+\alpha) \epsilon_{\alpha \beta} I_{-}, &&
\{ {\bar Q}^\a, {\bar S}^\b \}=-2i (1+\alpha) \epsilon^{\alpha \beta} I_{+},
\nonumber\\[2pt]
& [ D,Q_\a] = -\frac{1}{2} Q_\a\ , && [ D,S_\a] =\frac{1}{2} S_\a\ ,
\nonumber\\[2pt]
&
[ K,Q_\a ] =S_\a\ , && [ H,S_\a ]=-Q_\a\ ,
\nonumber\\[2pt]
&
[ \mathcal{J}_a,Q_\a] =\frac{i}{2} {{(\s_a)}_\a}^\b Q_\b\ , && [ \mathcal{J}_a,S_\a] =\frac{i}{2} {{(\s_a)}_\a}^\b S_\b\ ,
\nonumber\\[2pt]
& [ D,\bar Q^\a ] =-\frac{1}{2} \bar Q^\a\ , && [ D,\bar S^\a] =\frac{1}{2} \bar S^\a\ ,
\nonumber\\[2pt]
& [K,\bar Q^\a] =\bar S^\a\ , && [ H,\bar S^\a] =-\bar Q^\a\ ,
\nonumber\\[2pt]
&
[\mathcal{J}_a,\bar Q^\a] =-\frac{i}{2} \bar Q^\b {{(\s_a)}_\b}^\a\ , && [ \mathcal{J}_a,\bar S^\a] =-\frac{i}{2}
\bar S^\b {{(\s_a)}_\b}^\a\,
\nonumber\\[2pt]
&
[ I_{-},\bar Q^\a ] =\epsilon^{\alpha \beta} Q_\beta, && [ I_{-},\bar S^\a ] =\epsilon^{\alpha \beta} S_\beta,
\nonumber\\[2pt]
&
[I_{+},Q_\a] =-\epsilon_{\alpha\beta} \bar Q^\b, && [ I_{+},S_\a] =-\epsilon_{\alpha\beta} \bar S^\b,
\nonumber\\[2pt]
&
[I_3,Q_\a ] =\frac{i}{2} Q_\a, &&  [I_3,S_\a] =\frac{i}{2} S_\a,
\nonumber\\[2pt]
&
[ I_3,\bar Q^\a] =-\frac{i}{2} \bar Q^\a, &&  [ I_3,\bar S^\a ] =-\frac{i}{2} \bar S^\a,
\nonumber\\[2pt]
&
[ I_{-},I_3]=-i I_{-}, &&  [I_{+},I_3 ] =i I_{+},
\nonumber\\[2pt]
&
[ I_{-},I_{+}] =2 i I_{3},
\end{align}
where ${{(\s_a)}_\a}^\b$ are the Pauli matrices (for our conventions see Appendix A).
$H$, $K$ and $D$ are the generators of conformal subalgebra $so(2,1)$, $\mathcal{J}_a$ generate $su(2)$--subalgebra, while $I_{\pm}$, $I_3$ describe one more $su(2)$ in the Cartan basis. $Q_\a$ are supersymmetry  generators and $S_\a$ are their superconformal partners which obey the conjugation rules
\be\label{Hermitian_1}
Q_\a^\dag=\bar Q^\a, \qquad S_\a^\dag=\bar S^\a.
\ee
As is evident from the form of the superalgebra (\ref{algebra}), the conjugation rules for the remaining generators should be chosen in the form
\bea\label{Hermitian_2}
&&
H^\dag=-H, \quad D^\dag=-D , \quad K^\dag=-K, \quad \mathcal{J}_a^\dag=-\mathcal{J}_a, \quad
I_{\pm}^\dag=I_{\mp}, \quad I_3^\dag=-I_3.
\eea
For $\a=-1$ the superalgebra (\ref{algebra}) reduces to the semidirect sum of $su(1,1|2)$ and $su(2)$ algebras.

As the next step, we choose the subgroup $H=\{D, \mathcal{J}_3, I_{\pm}, I_3\}$ and construct the coset space $G/H$ where $G=D(2,1;\a)$. The reason for such a choice is to provide
the bosonic part of the coset space with the product structure $AdS_2 \times S^2$. Then one has to compute the Maurer-Cartan (MC) one--forms which are defined in the standard way
\bea\label{MC_1}
&&
\tilde G^{-1}d\tilde G=H L_H+K L_K+i\left(L_Q Q+\bar Q L_{\bar Q}+L_S S+\bar S L_{\bar S}\right)+\mathcal{J}_m L_m
\nonumber\\[2pt]
&&
\qquad \qquad   +D L_D +i(I_+L_{I_+}+I_-L_{I_-})+I_3L_{I_3}+ \mathcal{J}_3 L_J,
\eea
where $\tilde G$ is a coset space representative, $m=1,2$, and the explicit form of $L_H$, $L_K$ etc. is given below in Eq. (\ref{MC_Forms_Explicit}). The right hand side of this expression is in agreement with the conjugation properties (\ref{Hermitian_1}) and (\ref{Hermitian_2}) which render it antihermitian.

The right action of the supergroup $G$ on the coset space $G/H$ yields the coordinate transformation $Z\rightarrow Z'$.  As is known, MC one-forms $L^p$ on a given coset space transform homogeneously under the group action (see, e.g., the discussion in \cite{Ortin})
\be\label{MC_Forms_Transformation_General}
L^p(Z')=L^p(Z)+L^q(Z)\ve^C W_C^I(Z)f_{Iq}^p,
\ee
where $f_{Iq}^p$ are the structure constants of a superalgebra at hand, $\ve^C$ are the infinitesimal transformation parameters, $W_C^I$ is the so called $H$-compensator. Here the index $C$ encompases all the algebra generators, $I$ covers the subalgebra associated with $H$, while $p$ and $q$ label the coset space generators.
Using this formula for $D(2,1;\a)$ one finds the transformation rules for the bosonic MC one-forms on the coset space
\bea
&&
L_H\rightarrow L_H(1-\ve^CW_C^D), \quad
L_K\rightarrow L_K(1+\ve^CW_C^D),
\quad
L_m\rightarrow L_n(\delta_{mn}+\ve^CW_C^{J_3}\e_{3nm}).
\eea

Note that the presence of the fermionic generators does not affect the structure of the transformations which holds the same as in the case of the pure bosonic algebra.
Hence, the familiar quadratic forms
\be\label{QudraticForms}
L_HL_K, \qquad  L_mL_m,
\ee
are invariant under the group action. They are the building blocks for constructing in the kinetic term in a super 0-brane action.

A conventional means of building the Wess--Zumino (WZ) term in a super 0-brane action is to find a linear combination of external products of MC one-forms on the coset space which is an exact two--form.  Then one uses
the Stokes theorem to represent the WZ term as an integral over the one--dimensional space, which is usually called the reduced WZ term. Since the two--forms and exterior derivatives of MC forms are related by means of the Maurer--Cartan equations, one expects the reduced WZ term to be expressed via the linear combination of the original bosonic MC one--forms.  To find them, notice that the MC one-forms on the subgroup transform as connections (see, e.g., the discussion in \cite{Krivonos_lectures,Ortin}). In particular, the transformation rules for $L_J$ and $L_D$ under the left group action read
\be\label{ConnectionTranformation}
L_D\rightarrow L_D+df_D, \qquad L_J\rightarrow L_J+df_J,
\ee
where $f_D$ and $f_J$ are some functions whose explicit forms depend on the way in which one parametrizes the coset. Thus, one may use a linear combination of such one-forms in constructing the reduced WZ term.

Summarizing the discussion above, the super 0-brane action on the coset space of $D(2,1;\a)$ supergroup reads
\be\label{Action}
S=-\tilde m\int \sqrt{4 L_H L_K-c L_mL_m}-\int \left(a L_D+b L_J\right),
\ee
where $\tilde m$, $c$, $a$ and $b$ are constant parameters. Note that when all the fermionic variables are set to zero, the expression under the square root represents a metric on $AdS_2\times S^2$ with different curvature radii.

\vspace{0.5cm}

\noindent
{\bf 3. $\kappa$-symmetry}\\

\noindent
In order to ensure that (\ref{Action}) is invariant under the $\kappa$--symmetry, we follow the standard procedure (see, e.g., \cite{Metsaev_Tseytlin}).
Computing the exterior derivative of (\ref{MC_1}), one first obtains the Maurer--Cartan equations for the bosonic forms
\bea
&&
dL_H=-L_H\wedge L_D-2i L_Q\wedge L_{\bar Q},
\nonumber\\[2pt]
&&
dL_K=L_K\wedge L_D-2i L_S\wedge L_{\bar S},
\nonumber\\[2pt]
&&
dL_D=-2L_H\wedge L_K+2i\left(L_Q\wedge L_{\bar S}+L_S\wedge L_{\bar Q}\right),
\nonumber\\[2pt]
&&
dL_a=-\frac{1}{2}\e_{abc}L_b\wedge L_c+2\a\left(L_S\sigma_a\wedge L_{\bar Q}-L_Q\sigma_a\wedge L_{\bar S}\right).
\eea
Then one uses them to represent variations of MC one-forms in the form
\bea\label{MK_Variation}
&&
\d L_H=d [\d x_H] + [\d x_D] L_H-L_D[\d x_H]-2i \left([\d \psi]L_{\bar Q}-L_Q[\d \bar\psi]\right),
\nonumber\\[2pt]
&&
\d L_K=d [\d x_K] - [\d x_D] L_K+L_D[\d x_K]-2i \left([\d \eta]L_{\bar S}-L_S[\d \bar\eta]\right),
\nonumber\\[2pt]
&&
\d L_D=d [\d x_D] - 2[\d x_H] L_K+2[\d x_K] L_H+2i  \left([\d \psi]L_{\bar S}-L_Q[\d \bar\eta]+[\d \eta]L_{\bar Q}-L_S[\d \bar\psi]\right),
\nonumber\\[2pt]
&&
\d L_a= d[\d x_a]-\e_{abc}[\d x_b] L_c+2\a\left([\d \eta]\sigma_a L_{\bar Q}-L_S\sigma_a[\d \bar\psi]-[\d \psi]\sigma_a L_{\bar S}+L_Q\sigma_a[\d \bar\eta]\right),
\eea
where, following \cite{Anabalon_Zanelli}, we introduced the notation
\be
[\d Z^A]=\d Z^M L^A{}_M
\ee
for MC one-form $L^A=dZ^M L^A{}_M $. In particular,
$[\d x_H]$, $[\d x_K]$, $[\d x_D]$, $[\d x_a]$ are associate with the MC forms $L_H$, $L_K$, $L_D$, $L_a$, respectively, while $[\d \psi]$, $[\d \bar\psi]$ correspond to $L_Q$, $L_{\bar Q}$ and $[\d \eta]$, $[\d \bar\eta]$ refer to $L_S$, $L_{\bar S}$.

As is known \cite{Metsaev_Tseytlin}, the $\k$-symmetry transformations are characterized by vanishing of $[\d Z^A]$ which are related to the bosonic one--forms on the coset space, i.e.
\be\label{Kappa_Bosonic}
[\d x_H]=[\d x_K]=[\d x_m]=0.
\ee
Taking into account this criterion and using (\ref{MK_Variation}), one can write down a variation of the action (\ref{Action}) irrespective of a specific choice of the coset parametrization
 \bea\label{ActionVariation}
 &&
 \delta_\k S=2\int \left\{\frac{2i \tilde m L_H [\d \eta]-\a c \tilde m L_n [\d\psi]\sigma_n}{\sqrt{4L_HL_K-c L_mL_m}}-i[\d\psi](a+i\a b\sigma_3) \right\}L_{\bar S}
 \nonumber\\[2pt]
&&
+2\int \left\{\frac{2i \tilde m L_K [\d \psi]+\a c \tilde mL_n [\d\eta]\sigma_n}{\sqrt{4L_HL_K-c L_mL_m}}-i [\d\eta](a-i\a b\sigma_3) \right\}L_{\bar Q}+h.c.,
 \eea
where $\sigma_n$, $n=1,2$ and $\sigma_3$ designate the Pauli matrices and the boundary terms $d[\d x_D]$ and $d[\d x_J]$ were discarded. Comparing the terms involving $[\d \eta]$ and $[\d\psi]$, one concludes that the action is invariant under the $\kappa$--symmetry provided the restrictions
\be\label{ParametersRestriction}
c=\a^{-2},\qquad  \tilde m^2=a^2+(\a b)^2
\ee
hold.
For further convenience we choose a solution of the equation $\d_\k S=0$ in the following form:
\bea\label{KappaSymmetry}
&&
[\d \eta]=\kappa,  \qquad [\d \bar\eta]=\bar\kappa, \qquad  [\d \psi]=[\d \eta]\Omega , \qquad [\d \bar \psi]=\Omega^\dag[\d \bar\eta],
 \nonumber\\[2pt]
&&
 \Omega=\frac{\sqrt{4L_HL_K-\a^{-2}L_mL_m}}{2\tilde m L_K}(a-i\a b\sigma_3)+i\a^{-1}\sigma_m\frac{L_m}{2L_K},
 \eea
where $\k=\k^\a$ is an anticommuting infinitesimal parameter and $\bar\k_\a=(\k^\a)^\dag$ is its conjugate.

By construction, the action (\ref{Action}) is invariant under the $D(2,1;\a)$ global supersymmetry and the local $\kappa$--transformations provided
Eq. (\ref{ParametersRestriction}) holds. It also has the reparametrization invariance which implies that the number of bosonic dynamical degrees of freedom is three. In its turn, the $\kappa$-symmetry reduces the number of (real) fermionic degrees of freedom from eight to four, which thus correspond to the $(3,4,1)$--supermultiplet of $D(2,1;\a)$. Concluding this section, we notice that for $\a=-1$
our analysis reproduces that in \cite{Zhou,KZ}.

\vspace{0.5cm}

\noindent
{\bf 4. Gauge fixing and global supersymmetry}\\

\noindent
The discussion above was independent of a specific parametrization of the coset. To be more precise, let us choose the following parametrization:
\be\label{CosetParametrization}
\tilde G=e^{tH}e^{zK}e^{i(\psi Q+\bar Q\bar\psi)}e^{i(\eta S+\bar S\bar\eta)}\tilde G_R,
\ee
where $\tilde G_R$ designate a coset representative of the rotation subgroup
\be
\tilde G_R=e^{\phi \mathcal{J}_1}e^{(\t-\pi/2)\mathcal{J}_2}.
\ee
Using this parametrization of the coset one can compute MC one-forms and the corresponding action (\ref{Action}).

In general, the $\kappa$--symmetry reduces the number of the fermionic degrees of freedom by half. For the case at hand this is achieved by imposing the gauge fixing condition
\be\label{Gauge_Fixing}
\eta=\bar\eta=0.
\ee
Note that usually there is a subtlety in gauge fixing the $\k$--symmetry in a way consistent with static solutions \cite{Sorokin} (see also \cite{KZ})~\footnote{We thank Dmitri Sorokin for pointing this out to us.}. However, it can be directly verified that (\ref{Gauge_Fixing}) is compatible with the static solution and, as thus, can be used without loss of the generality.

In order to simplify the analysis, in what follows we deal with the gauge fixed objects only.
The gauge fixed MC one-forms read
\begin{align}\label{MC_Forms_Explicit}
&
L_H=dt-i( \psi d\bar \psi-d\psi \bar\psi)-(1+2\a)L_K(\psi\bar \psi)^2, && L_K=z^2dt+dz,
\nonumber\\[2pt]
&
L_a=L^0_a-2\a L_K(\psi\sigma_b\bar \psi)R_{ba}, && L_D=2zdt
\nonumber\\[2pt]
&
L_Q=\left(d\psi+i(1+2\a)L_K(\psi\bar \psi)\psi-zdt\psi\right)\Gamma, && L_S=L_K\psi\Gamma,
\nonumber\\[2pt]
&
L_{\bar Q}=\Gamma^\dag\left(d\bar \psi-i(1+2\a)L_K(\psi\bar \psi)\bar \psi-zdt\bar \psi\right), && L_{\bar S}=L_K\Gamma^\dag\bar \psi,
\nonumber\\[2pt]
&
L_{I_{+}}=-i(1+\a)L_K\bar\psi^2, && L_{I_{-}}=i(1+\a)L_K\psi^2,
\nonumber\\[2pt]
&
 L_{I_3}=2(1+\a)L_K (\psi\bar\psi),
\end{align}
with
\be
L^0_1=\sin\t d\phi, \qquad L^0_2=d\t, \qquad L^0_3=-\cos\t d\phi.
\ee
Above we used the notation for the coset space element $\tilde G_R$ corresponding to the rotations subgroup in the adjoint $R_{ba}$ and the spinorial $\Gamma$ representations
\be
\tilde G_R^{-1}\mathcal{J}_a\tilde G_R=R_{ab}\mathcal{J}_b, \qquad \tilde G_R^{-1}Q\tilde G_R=\Gamma Q,
\ee
which have the explicit form
\bea
&&
R_{ab}=\left(
\begin{array}{cccccc}
\sin\t & 0 & -\cos\t  \\
\nonumber\\[1pt]
-\cos\t\sin\phi & \cos\phi & -\sin\t\sin\phi\\
\nonumber\\[1pt]
\cos\t\cos\phi & \sin\phi & \sin\t\cos\phi
\end{array}
\right), \qquad
\Gamma=e^{-\frac{i}{2}\phi\sigma_1}e^{-\frac{i}{2}\left(\t-\frac{\pi}{2}\right)\sigma_2}
\eea
and obey the relations
\be\label{Bilinear}
\G\sigma_a\G^\dag=R_{ba}\sigma_b, \qquad R_{ab}R_{ac}=\d_{bc}.
\ee

The left action of the supergroup $D(2,1;\a)$ on the coset space (\ref{CosetParametrization}) yields global symmetry transformations of the action (\ref{Action}) before gauge fixing. It turns out that the gauge condition (\ref{Gauge_Fixing}) is compatible with the bosonic symmetries only. After fixing the gauge, the coordinate transformations have the following form:
\bea\label{BosonicGenerators}
&&
\d_H t=1,
\nonumber\\[2pt]
&&
\d_D t=-t, \quad \d_D z=z, \quad \d_D \psi=-\frac12\psi, \quad \d_D \psi=-\frac12\bar\psi,
\nonumber\\[2pt]
&&
\d_K t=t^2, \quad \d_K z=1-2tz, \quad \d_K \psi=t\psi, \quad \d_K \psi=t\bar\psi,
\nonumber\\[2pt]
&&
\d_a\phi=\sin^{-1}\t R_{a1}, \quad  \d_a\t= R_{a2}, \quad \d_a\psi=\frac{i}{2}\psi\sigma_a, \quad \d_a\bar\psi=-\frac{i}{2}\sigma_a\bar\psi,
\nonumber\\[2pt]
&&
\d_{I_3}\psi=\frac{i}{2}\psi, \qquad \d_{I_3}\bar\psi=-\frac{i}{2}\bar\psi, \qquad \d_{I_+}\bar\psi=-\psi, \qquad \d_{I_-}\psi=\bar\psi,
\eea
where $H$, $D$, $K$, $\mathcal{J}_a$, $I_{\pm}$, $I_3$ stand for time translations, dilatations, special conformal transformations, rotations,  and $su(2)$--transformations, respectively. For simplicity, we set infinitesimal parameters to be equal to unity. It is straightforward to verify that the action (\ref{Action}) constructed in terms of gauged fixed MC one-forms (\ref{MC_Forms_Explicit}) is invariant under the transformations (\ref{BosonicGenerators}).

The gauge fixing condition (\ref{Gauge_Fixing}) is not invariant under the supersymmetry and superconformal transformations (for more details see Appendix B). The conventional means of restoring such symmetries is to implement a compensating $\kappa$--transformation. For what follows it proves convenient to redefine the temporal and fermionic coordinates
  \be\label{Redefinition}
  t\rightarrow t+\frac{1}{z}, \qquad \psi\rightarrow \frac{\psi}{z},\qquad \bar\psi\rightarrow \frac{\bar\psi}{z}.
  \ee
which yields the supersymmetry transformations (see also Appendix B)
\bea\label{Q-Transformations_2}
&&
\d_Q z=-(\e\bar\psi)(\psi\bar\psi)-iz(\e\bar\psi),
\nonumber\\[2pt]
&&
\d_Q \phi=2\a\sin^{-1}\t(\e\sigma_a\bar\psi)R_{a1}, \qquad \d_Q \t=2\a(\e\sigma_a\bar\psi)R_{a2},
\nonumber\\[2pt]
&&
\d_Q\psi=\e\left(z+i(\psi\bar\psi)\right)+i(1+\a)(\e\bar\psi)\psi, \qquad \d_Q\bar\psi=-i(1-\a)(\e\bar\psi)\bar\psi,
\nonumber\\[2pt]
&&
\d_{\tilde\k} z=i z^2\left(\e \G^\dag\O\G\bar\psi\right), \qquad \d_{\tilde\k}\psi=z^2\left(\e \G^\dag\O\G\right),
\eea
and their superconformal partners
\bea\label{S-Transformations}
&&
\d_S z=-i(1+tz)(\e\bar\psi)-t(\e\bar\psi)(\psi\bar\psi),
\nonumber\\[2pt]
&&
\d_Q \phi=2\a t\sin^{-1}\t(\e\sigma_a\bar\psi)R_{a1}, \qquad \d_Q \t=2\a t(\e\sigma_a\bar\psi)R_{a2},
\nonumber\\[2pt]
&&
\d_S \psi=\e(1+tz)+itz\left(\e(\psi\bar\psi)-(1+\a)\psi\right), \qquad \d_S \bar\psi=-i(1-\a)t(\e\bar\psi)\bar\psi,
\nonumber\\[2pt]
&&
\d_{\tilde\k}z=itz^2\left(\e \G^\dag\O\G\bar\psi\right), \qquad \d_{\tilde\k}\psi=t z^2\left(\e \G^\dag\O\G\right),
\eea
where $\O$ is given in Eq. (\ref{KappaSymmetry}) above and similar transformations $\d_{\bar Q}$, $\d_{\bar S}$ are obtained by hermitian conjugation.

\vspace{0.5cm}

\newpage
\noindent
{\bf 5. Bosonic part of the action}\\

\noindent
Before analyzing the full supersymmetric gauge fixed action (\ref{Action}), let us consider its bosonic part and demonstrate that it can be linked to a massive charged particle propagating near the horizon of the extreme Reissner--Nordstr\"om--AdS--dS black hole. Redefining the coordinates in MC one-forms  (\ref{MC_Forms_Explicit}) as follows:
  \be
  t\rightarrow \frac12 \left(t+\frac{1}{r}\right), \qquad z\rightarrow r,
  \ee
  one rewrites the action (\ref{Action}) in the conventional $AdS$ basis
\be\label{Action_Bosonic}
S=-\tilde m\int dt \left(r^2-\dot{r}^2/r^2-\a^{-2}(\dot\t^2+\sin^2\t\dot\phi^2)\right)^{1/2}-\int dt \left(a r -b \cos\t\dot\phi\right),
\ee
up to the total derivative. On the other hand, the action functional of a particle probe propagating near the horizon of the extreme Reissner--Nordstr\"om--AdS--dS black hole reads (see, e.g., \cite{Romans})
  \bea\label{ReissnerNordstrom}
  &&
  S=-m \left(\frac{2}{V''(r_+)}\right)^{1/2}\int \left(r^2-\frac{\dot{r}^2}{r^2}-V''(r_+)\frac{r_+^2}{2}(\dot\t^2+\sin^2\t\dot\phi^2)\right)^{1/2}
  \nonumber\\[2pt]
&&
 \qquad \qquad \qquad \qquad \qquad \qquad \qquad \qquad \qquad  -q\int \left(\frac{2Qr}{r_+^2 V''(r_+)}-P\cos\t\dot\phi\right),
  \eea
 where $Q$ and $P$ are the electric and magnetic charges of the black hole, while $m$ and $q$ are the mass and electric charge of a test particle. In the previous formula $V(r)$ is given by
 \be
  V=1-\frac{2 M}{r}+\frac{Q^2+P^2}{r^2}+\frac{1}{3}\L r^2,
 \ee
 where $M$ is the black hole mass and $\L$ is a cosmological constant, and $r_+$ refers to the horizon radius
 \be
 V(r_+)=V'(r_+)=0 \quad \rightarrow \quad Q^2+P^2=r_+^2(1+\L r_+^2), \qquad M=r_+\left(1+\frac{1}{3}\L r_+^2\right).
 \ee

One thus concludes that (\ref{Action_Bosonic}) coincides with (\ref{ReissnerNordstrom}) provided the identification
\be
\tilde m=m \left(\frac{2}{V''(r_+)}\right)^{1/2}, \qquad a=\frac{2Qq}{r_+^2 V''(r_+)},\qquad b=q P, \qquad \a^{-2}=\frac{r_+^2}{2}V''(r_+)
\ee
holds, while the restriction  (\ref{ParametersRestriction}) takes the form
\be\label{ParametersRestriction_1}
m^2=\frac{2q^2 Q^2}{r_+^4 V''(r_+)}+\frac{q^2 P^2}{r_+^2}.
\ee
Note that the Reissner--Nordstr\"om--AdS--dS black hole can be viewed as a supergravity solution only for the vanishing magnetic charge $P=0$ and the negative cosmological constant  \cite{Romans, Klemm}. Thus, while for $\Lambda=0$ the test particle propagates on a $BPS$ background, it fails to do so for $\Lambda\ne 0$.

\vspace{0.5cm}

\noindent
{\bf 6. Hamiltonian formulation}\\

\noindent
Having discussed the bosonic part of the action, let us turn to its supersymmetric extension. Our aim in this section is to argue that the super 0--brane (\ref{Action}) in the gauge fixed form can be
linked to the $D(2,1;\alpha)$--superparticle recently constructed in \cite{Galajinsky_D} within the Hamiltonian framework. Below we demonstrate that the two models are
related by a canonical transformation.

After implementing the change of coordinates (\ref{Redefinition}), the gauge fixed supersymmetric action reads
\bea\label{Action_Explicit}
&&
S=-\tilde m\int\Big[4(z^2-\dot z)-\a^{-2}(\dot\t^2+\sin^2\t\dot{\phi}^2)-4i (\psi\dot{\bar{\psi}}-\dot\psi\bar{\psi})-4\a^{-1}L^0_m R_{bm} (\psi\sigma_b\bar\psi)
\nonumber\\[2pt]
&&
\qquad \qquad-4(2\a-1)(\psi\bar\psi)^2\Big]^{1/2}dt -\int\left(2 a z -b \cos\t\dot\phi-2\a b(\psi\sigma_a\bar\psi)R_{a3}\right)dt.
\eea
When obtaining the last formula the property of the rotation matrix (\ref{Bilinear}) and the spinor identity
\be
(\psi\sigma_a\bar\psi)(\psi\sigma_b\bar\psi)=-\d_{ab}(\psi\bar\psi)^2,
\ee
prove to be helpful. The advantage of the coordinate redefinition (\ref{Redefinition}) is that the expression under the square root in (\ref{Action_Explicit}) is linear in $\dot z$ and it does not involve cross terms with the fermionic velocities.

Introducing the momenta $(p_z, p_\t, p_\phi)$ canonically conjugate to the bosonic variables $(z,\t,\phi)$, one finds the Hamiltonian
\be\label{HH}
H=z^2p_z+\frac{a^2}{p_z}+2 a z+\frac{\a^2}{p_z}J_aJ_a+2\a(\psi\sigma_a\bar\psi)J_a-p_z(1+2\a)(\psi\bar\psi)^2,
\ee
where
\bea\label{Generators_Rotation}
&&
J_1=p_\phi, \qquad J_2=p_\t\cos\phi-p_\phi\cot\t \sin\phi+b\frac{\sin\phi}{\sin\t},
\nonumber\\[2pt]
&&
J_3=p_\t\sin\phi+p_\phi\cot\t \cos\phi-b\frac{\cos\phi}{\sin\t}.
\eea
Momenta $(p_\psi, p_{\bar\psi})$ canonically conjugate to the fermionic variables $(\psi, \bar\psi)$ give rise to the second class constraints
\be\label{Constraints}
p_{\psi}-i\bar\psi p_z=0, \qquad p_{\bar\psi}-i\psi p_z=0,
\ee
where we have chosen the right derivative for the fermionic variables.

In order to construct a canonical transformation which links (\ref{HH}) to the Hamiltonian in \cite{Galajinsky_D},
let us change the fermions
\be\label{Psi_Redefinition}
\psi\rightarrow \frac{\psi}{\sqrt{2p_z}}, \qquad \bar\psi\rightarrow \frac{\bar\psi}{\sqrt{2p_z}}, \qquad p_\psi\rightarrow \sqrt{2p_z}p_\psi, \qquad  p_{\bar\psi}\rightarrow \sqrt{2p_z}p_{\bar\psi}.
\ee
It is easy to see that this transformation is canonical since the corresponding compensating transformation
  \be
  z\rightarrow z+ \frac{1}{2p_z}(\psi p_\psi+\bar\psi p_{\bar\psi}),
  \ee
 is identical on the constraint surface (\ref{Constraints}). Finally, let us canonically redefine the bosonic coordinate $z$ and its conjugate momentum $p_z$
\be\label{CanonicalTransformation}
z\rightarrow -\frac{p}{x}-\frac{2 a}{x^2}, \qquad p_z\rightarrow \frac{x^2}{2}.
\ee
This transformation along with (\ref{Psi_Redefinition}) brings the Hamiltonian and the constraint surface to those in \cite{Galajinsky_D}. As a by--product at $\a=-1$ one obtains a canonical transformation which links the $SU(1,1|2)$--invariant models in \cite{Zhou} and \cite{bgik,g1}.

\vspace{0.5cm}

\noindent
{\bf 7. Conclusion}\\

\noindent
To summarize, in this work we have constructed the super 0-brane action on the coset space of the $D(2,1;\a)$ supergroup. To this end the method of nonlinear realizations was used. First we defined the coset superspace and investigated MC one--forms which provided the building blocks for constructing the action. Analyzing $\kappa$--invariance we have found a restriction which linked free parameters entering the action. It generalized the condition obtained in \cite{Zhou} for the super 0-brane on the coset space of the $SU(1,1|2)$ supergroup. The gauge fixing procedure and the global supersymmetry transformations of the gauge fixed action were described. Focusing on the bosonic part of the action, we have shown that it describes a massive charged particle propagating near the horizon of the extreme Reissner--Nordstr\"om--AdS--dS black hole.
In contrast to the case of a vanishing cosmological constant \cite{Zhou}, in general the super 0-brane propagate on a non--$BPS$ background. The gauge fixed form of the super 0--brane was then
linked to the $D(2,1;\alpha)$--superparticle in \cite{Galajinsky_D} by constructing a suitable canonical transformation within the Hamiltonian formalism. At $\a=-1$ the transformation relates the $SU(1,1|2)$--invariant models in \cite{Zhou} and \cite{bgik,g1}.

\vspace{0.5cm}

\noindent
{\bf Acknowledgements}\\

\noindent
I would like to thank  A. Galajinsky for posing the problem, helpful discussions, and reading the manuscript. I am also grateful to Igor Bandos and especially to Dmitri Sorokin for useful discussion and valuable comments. This work was supported by the RF Presidential grant MK-2101.2017.2 and
the Tomsk Polytechnic University competitiveness enhancement program, project CEP-PTI-72/2017.

\vspace{0.5cm}
\noindent
{\bf  Appendix A. Spinor notations}\\

In this work, we use the notation in which spinor indices are raised and lowered with the use of the Levi--Cevita symbol
\be
\psi_\a=\ve_{\a\b}\psi^\b, \qquad \bar\psi^\a=\ve^{\a\b}\bar\psi_\b,
\nonumber
\ee
where $\ve_{12}=1$, $\ve^{12}=-1$ and $\bar\psi_\a=(\psi^\a)^*$. We often skip the spinor indices on the fermionic variables
 \be
 \psi=\psi^\a, \qquad \bar\psi=\bar\psi_\a,
 \nonumber
 \ee
and assume the summation over repeated indices
\be
\psi\bar\psi=\psi^\a\bar\psi_\a, \qquad \psi Q=\psi^\a Q_\a, \qquad \psi\sigma_\a\bar\psi=\psi^\a(\sigma_a)_\a{}^\b\bar\psi_\b, \qquad \psi^2=\psi^\a \psi_\a, \qquad \bar\psi^2=\bar\psi_\a \bar\psi^\a.
\nonumber
\ee
The Pauli matrices are chosen in the standard form
 \be
\s_1=\begin{pmatrix}0 & 1\\
1 & 0
\end{pmatrix}\ , \qquad \s_2=\begin{pmatrix}0 & -i\\
i & 0
\end{pmatrix}\ ,\qquad
\s_3=\begin{pmatrix}1 & 0\\
0 & -1
\end{pmatrix}.
\nonumber
\ee
They obey the properties
\bea
&&
\{\s_a,\s_b\}=2\d_{ab}, ~
[\s_a,\s_b]=2i \e_{abc} \s_c, ~
\s_a \s_b=\d_{ab} +i \e_{abc} \s_c, ~ {{(\s_a)}_\a}^\b {{(\s_a)}_\g}^\r=2 {\d_\a}^\r {\d_\g}^\b-{\d_\a}^\b {\d_\g}^\r,
\nonumber
\eea
which are extensively used in the text.

\vspace{0.5cm}

\noindent
{\bf  Appendix B. Supersymmetry transformations}\\

\noindent
In this appendix we display the explicit form of the global supersymmetry transformations.

Consider the left action of the group element $e^{\e Q}$ on the coset space (\ref{CosetParametrization}) with $\eta=\bar\eta=0$
\be\label{Q_Action}
e^{\e Q} \tilde G|_{\eta=\bar\eta=0}=e^{H(t+\d_Q t)}e^{K(z+\d_Q z)}e^{iQ(\psi+\d_Q \psi)+i(\bar\psi+\d_Q\bar\psi)\bar Q} e^{-iz (\e S)}\tilde G_R' h,
\nonumber
\ee
 where
\bea
&&
 \tilde G_R'=e^{J_1(\phi+\d_Q \phi)}e^{J_2(\t+\d_Q\t+\pi /2)},
\nonumber\\[2pt]
&&
h=e^{-2zD\left(\d_Q t-i(\e\bar\psi)\right)} e^{2z \sin^{-1}\t \left((\e\s_2\bar\psi)\sin\phi-(\e\s_3\bar\psi)\cos\phi\right) J_3}e^{-2(1+\a)(\e\bar\psi)I_3-2i(1+\a)\psi^\a\e^\b\ve_{\a\b}I_-},
\nonumber\\[2pt]
&&
\d_Q t=z (\e\bar\psi)(\psi\bar\psi)+i(\e\bar\psi), \qquad \d_Q z=-z^2\d_Q t,
\nonumber\\[2pt]
&&
\d_Q \phi=2\a z\sin^{-1}\t(\e\sigma_a\bar\psi)S_{a1}, \qquad \d_Q \t=2\a z(\e\sigma_a\bar\psi)S_{a2},
\nonumber\\[2pt]
&&
\d_Q\psi=\e\left(1+iz(\psi\bar\psi)\right)+i(1+\a)z(\e\psi)\psi+z\psi\d_Q t, \qquad \d_Q\bar\psi=i\a z\bar\psi(\e\bar\psi),
\nonumber
\eea
and $\e$ is an infinitesimal Grassmann-valued parameter of the transformation. From (\ref{Q_Action}) one obtains the transformation rules for the fermionic coordinates
 \be
 \d_Q\eta|_{\eta=\bar\eta=0}=-\e z, \qquad \d_Q\bar\eta|_{\eta=\bar\eta=0}=0.
 \nonumber
 \ee
The leftmost equation implies that the gauge fixing condition (\ref{Gauge_Fixing}) is not invariant under the supersymmetry transformations.
Compensating  transformation $\d_{\tilde\k}$ should satisfy the condition
\be
(\d_{\tilde\k}+\d_Q)\eta=(\d_{\tilde\k}+\d_Q)\bar\eta=0,
\ee
which implies
\be
\d_{\tilde\k}\eta=z\e, \qquad  \d_{\tilde\k}\bar\eta=0.
\ee
In order to find compensating transformations for the bosonic $t$, $z$, $\t$, $\phi$ and fermionic $\psi$, $\bar\psi$ coordinates, one has to treat  (\ref{Kappa_Bosonic}) and (\ref{KappaSymmetry}) as a system of algebraic equations in $\d_{\tilde \k}t$, $\d_{\tilde \k}z$, $\d_{\tilde \k}\t$, $\d_{\tilde \k}\phi$ and $\d_{\tilde \k}\psi$, $\d_{\tilde \k}\bar\psi$ with MC one-forms given in (\ref{MC_Forms_Explicit})
    \bea
&&
[\d\eta]=z \e \G, \qquad [\d\bar\eta]=0,
\nonumber\\[2pt]
&&
[\d\psi]=\left(\d_{\tilde\k}\psi-z\d_{\tilde\k}t\psi\right)\Gamma=[\d\eta] \Omega,
\nonumber\\[2pt]
&&
[\d\bar\psi]=\G^{\dag}\left(\d_{\tilde\k}\psi-z\d_{\tilde\k}t\bar\psi\right)=\Omega^\dag [\d\bar\eta],
\nonumber\\[2pt]
&&
[\d x_H]=\d_{\tilde\k}t-i( \psi \d_{\tilde\k}\bar \psi-\d_{\tilde\k}\psi \bar\psi)=0, \qquad [\d x_K]=z^2\d_{\tilde\k}t+\d_{\tilde\k}z=0,
\nonumber\\[2pt]
&&
[\d x_1]=\sin\t \d_{\tilde\k}\phi=0, \qquad [\d x_2]=\d_{\tilde\k}\t=0.
\nonumber
\eea
The general solution of this system is given by
 \bea\label{Q_Compensating_Transformation}
&&
\d_{\tilde\k}t=-iz \left(\e \G\Omega\G^\dag\bar\psi\right), \qquad \d_{\tilde\k}z=-z^2\d_{\tilde\k}t, \qquad \d_{\tilde\k}\phi=\d_{\tilde\k}\t=0,
\nonumber\\[2pt]
&&
\d_{\tilde\k}\psi=z\left(\e\G\O\G^\dag\right)+z\d_{\tilde\k}t \psi, \qquad \d_{\tilde\k}\bar\psi=z \d_{\tilde\k}t \bar\psi,
\nonumber
\eea
where $\G\O\G^\dag$ can be written in the form
\be
\G\O\G^\dag=\frac{\sqrt{4L_HL_K-\a^{-2}L_mL_m}}{2m L_K}\left(a+i b\sigma_a R_{a3}\right)-i\a^{-1}\sigma_a R_{am}\frac{L_m}{2L_K}.
\nonumber
\ee
In the same way one may find the action of the superconformal generator $S$ on the coset (\ref{CosetParametrization})
\bea\label{S_transformations_1}
&&
\d_S t=-it(\e\bar\psi)-(1-tz)(\e\bar\psi)(\psi\bar\psi), \qquad \d_S z=-z^2\d_S t,
\nonumber\\[2pt]
&&
\d_S\psi=i(1-tz)\left((\psi\bar\psi)\e-(1+\a)\psi\right)+t\e+z\psi \d_S t, \qquad \d_S \bar\psi=-i(1-\a)(1-tz)(\e\bar\psi)\bar\psi,
\nonumber\\[2pt]
&&
\d_S\phi=2\a(1-tz)\sin^{-1}\t R_{a1}(\e\sigma_a\bar\psi), \qquad \d_S\t=2\a(1-tz)R_{a2}(\e\sigma_a\bar\psi)
\nonumber
\eea
and the corresponding compensating transformations
\bea\label{S_CompensatingTransformations}
&&
\d_{\tilde\k}t=i(1-t z)\left(\e \G\Omega\G^\dag\bar\psi\right), \qquad \d_{\tilde\k}z=-z^2\d_{\tilde\k}t,
\nonumber\\[2pt]
&&
\d_{\tilde\k}\psi=-(1-tz)\left(\e\G\O\G^\dag\right)+z\d_{\tilde\k}t \psi, \qquad \d_{\tilde\k}\bar\psi=z \d_{\tilde\k}t \bar\psi.
\nonumber
\eea
The $\bar Q$--, and $\bar S$--transformations follow by hermitian conjugation.

\end{document}